\documentclass{aa}

\usepackage{hyperref}
\usepackage{xcolor}
\usepackage{lscape}

\usepackage{graphicx}
\newcommand{\kms}{km\,${\rm s}^{-1}$}

\usepackage{txfonts}

\begin{document}

   \title{Investigating the lack of main-sequence companions to massive Be stars}

   \author{J. Bodensteiner,
   T.~Shenar,
   H.~Sana
   }

   \institute{Instituut voor Sterrenkunde, KU Leuven, Celestijnenlaan 200D, 3001 Leuven, Belgium\\
              \email{julia.bodensteiner@kuleuven.be}
    }

   \date{Received 2 February 2020; accepted 17 June 2020}


  \abstract
   {About 20\% of all B-type stars are classical Be stars - stars whose spectra imply the presence of a circumstellar decretion disk. The disk phenomenon is strongly correlated with rapid rotation, the origin of which remains unclear. It may be rooted in single- or binary-star evolution. In the framework of the binary channel, the initially more massive star transfers mass and angular momentum to the original secondary, which becomes a Be star. The system then evolves into a Be binary with a post-main-sequence companion, which, depending on the companion mass, may later be disrupted in a supernova event. Hence, if the binary channel dominates the formation of Be stars, one may expect a strong lack of close Be binaries with main sequence (MS) companions.}
   {We want to test the prediction of the binary channel. Through an extensive, star-by-star review of the literature of a magnitude-limited sample of Galactic early-type Be stars, we investigate whether Be binaries with MS companions are known to exist.}
   {Our sample is constructed from the BeSS database and cross-matched with all available literature on the individual stars. Archival and amateur spectra are used to verify the existing literature when conflicting reports are found.}
   {Out of an initial list of 505 Be stars, we compile a final sample of 287 Galactic Be stars earlier than B1.5 with $V\leq12\,$mag. Out of those, 13 objects were reported as Be binaries with known post-MS companions (i.e., compact objects or helium stars) and 11 as binaries with unknown, uncertain or debated companions. We find no confirmed reports of Be binaries with MS companions. For the remaining 263 targets, no significant reports of multiplicity exist in the literature, implying that they are either Be binaries with faint companions, or truly single.}
   {The clear lack of reported MS companions to Be stars, which stands in contrast to the high number of detected B+B MS binaries, strongly supports the hypothesis that early-type Be stars are binary interaction products that spun up after mass and angular momentum transfer from a companion star. Taken at face value, our results may suggest that a large majority of the early-type Be stars have formed through binary mass-transfer. } 
   \keywords{stars: massive, early-type, emission-line, Be - binaries: spectroscopic, close}

   \authorrunning{Bodensteiner et al.}

   \maketitle
%

%
\section{Introduction}\label{Sec:intro}
About 20\% of the Galactic non-supergiant B-type stars exhibit Balmer emission lines
(most prominently in H$\alpha$), defining the class of classical Be stars
\citep[see e.g.,][]{Zorec1997, Rivinius2013}. This emission is thought to arise from a circumstellar decretion disk that extends several stellar radii along the equatorial plane of the star \citep{Struve1931, Gies2007, Carciofi2009}. The Be phenomenon is known to be transient, with emission lines appearing and disappearing over the timescales of months, years, or decades \citep{Rivinius2013}.

While the formation of the disk is not fully understood, there is a consensus that its presence strongly correlates with rapid rotation of the Be star \citep[see e.g.,][]{Porter2003}. Significant rotation acts to reduce the equatorial escape velocity. This is a fundamental prerequisite for the majority of disk-formation models, which invoke mechanisms such as turbulence  \citep[e.g.,][]{Townsend2004} or non-radial pulsations \citep{Baade1988, Baade2017, Semaan2018}
for feeding the disk.
Even though it is unclear whether all Be stars rotate near-critically or not, the observed distribution of their projected rotational velocities ($\varv \sin i$) implies that virtually all Be stars are rapid rotators \citep{Townsend2004, Zorec2016}. A central question in the context of Be-star formation is therefore the origin of the rapid stellar spins. In this work, we focus on the origin of rapid rotation in Be stars, for which three main alternatives have been proposed.

Two of the commonly discussed mechanisms are rooted in single-star evolution. On the one hand, Be stars could be born as rapid rotators, having inherited the angular momentum from their parental molecular cloud \citep{Bodenheimer1995}. This, however, is contradicted by observations: Be stars are found in clusters of all ages \citep{Abt1979, Mermilliod1982, Slettebak1985}, while they seem to reach a maximum abundance in clusters between 13 and 25 Myr \citep{Fabregat2000, Tarasov2017}. Furthermore, the rotation rates found for young B-type stars seem to be below the limit for Be star formation \citep{Huang2010}.

On the other hand, B stars may experience a spin up during their main-sequence (MS) evolution via angular momentum transfer from the core to the envelope \citep{Langer1998, Meynet2000, Ekstrom2008, Granada2013}. It was recently demonstrated by \citet{Hastings2020} that this mechanism can indeed account for near-critical rotation towards the end of the MS, when the stellar envelope expands. However, as the authors discuss, the model encounters several problems when confronted with observations \citep[e.g., the lack of observed nitrogen enhancement, reported for example by][]{Ahmed2017}. Moreover, by adopting the observed distribution of rotational velocities of presumably single B stars \citep{Dufton2013} as initial rotations in their models, Hastings et al. have likely overestimated the amount of rotation. The reason is that given the star formation history of the sample \citep{Schneider2018}, many of the rapidly-rotating stars are potentially binary-interaction products \citep[see e.g.,][]{deMink2014}.

Alternatively, Be stars could have gained their rapid rotation through mass and angular momentum transfer in binary interactions \citep{Kriz1975, Pols1991, Langer2019}. In this case, the observed Be star was originally the secondary component (i.e., the initially less massive component), having accreted mass from the Roche-lobe overfilling primary component. If the original primary (i.e., the mass donor) avoided merging with the Be progenitor during mass-transfer, it is now either a He-burning stripped star such as an O- or B-type subdwarf (sdO, sdB) or a Wolf-Rayet (WR) star, or a compact object such as a white dwarf (WD), neutron star (NS), or black hole (BH). If the mass donor ends its life in a supernova, the system may be disrupted, forming a single Be star with a history of binary interaction \citep{Blaauw1961, Gies1986}. Chances are that such companions will elude detection \citep{deMink2014}. Based on detailed binary evolution calculations,  \citet{Shao2014} report that most Be stars may be the products of binary interactions. Binary population synthesis computations, assuming a single-starburst, predict the fraction of massive binary interaction products to peak at a cluster age of $\sim$8-20\,Myr \citep{Schneider2015}, i.e. in line with the observations quoted before.

It remains unclear whether the origin of rapid rotation in Be stars is rooted primarily in one of the above-mentioned mechanisms, or in a combination thereof.
Considering that the majority of massive stars will interact with a companion during their lifetime \citep{Sana2012, Dunstall2015}, it is certainly conceivable that the binary channel is responsible for the formation of the vast majority - if not all - of the massive Be stars.

Different types of Be star systems with an evolved companion are known. The systems with the most massive companions are Be/X-ray binaries \citep[BeXRBs, see e.g.,][]{Reig2011}. While the nature of the companion is still unknown for a significant fraction of BeXRBs, most known companions are neutron stars. There is currently only one Be star known with a BH companion \citep{Casares2014}. Using observations in the ultraviolet from the International Ultraviolet Explorer (IUE) satellite, a handful of Be binaries with sdB or sdO companions were confirmed, including the well-known case of $\varphi$ Per \citep{Peters2008, Peters2013, Peters2016, Wang2017, Wang2018}.

Several authors have studied the binary origin of Be stars before. \citet{Abt1978} studied the binary fraction of a sample of almost 60 galactic Be stars of spectral type B2 to B5 through multi-epoch spectroscopy. Based on low number statistics, they find similar binary fractions for B and Be stars. They find, however, a significant difference in the period distribution: while half of the B star binaries have periods below 100 days, there is no Be binary with such periods. Furthermore, taking a closer look at their Be binaries shows that none of them, taking recent literature into account, is a double-lined spectroscopic binary (SB2).

\citet{Oudmaijer2010} investigated the binary fraction of Be stars in comparison to B stars through high-angular-resolution imaging. They found a similar binary fraction for B and Be stars, and concluded that the binary channel is probably not responsible for the formation of Be stars. However, their survey was only sensitive to very long orbital periods of at least ${\approx}5000\,$d, while post-interaction Be binaries are expected to exhibit much shorter periods of the order of one\,year \citep[e.g.,][]{Langer2019}. Similarly, the recent speckle imaging survey performed by \cite{Horch2020} does not probe systems that are tight enough for binary evolution to have played a role. 

In contrast, \citet{McSwain2005} examined the occurrence of the Be phenomenon as a function of stellar age and evolutionary stage through a photometric study of southern open clusters. They found that the fraction of Be stars is higher among earlier type and slightly evolved stars \citep[see also][]{Zorec1997, Martayan2010}. With this, the authors tested different theories proposed for the origin of rapid rotation in Be stars. They argue that a higher fraction of late-type Be stars would be expected if Be stars were born as rapid rotators or if they spin-up at the terminal-age MS, and concluded that their observed spectral type distribution is consistent with Be stars being spun-up by binary mass transfer. \citet{Shokry2018}, on the other hand, found observational indications that there is no strong dependence of the Be star fraction on the spectral type because the fraction among late-type stars is underestimated.

\citet{Berger2001} studied the kinematics of $\sim$350 Be stars using Hipparcos proper motions and published radial velocities (RVs). They report that 3-7\% of the Be stars in their sample are runaway stars and interpret this as an indication that a fraction of them have formed in binary system. More recently, \citet{Boubert2018} performed a similar study of $\sim$650 Be stars based on the first release of GAIA data. They found that the observed fraction of runaway Be stars in their sample is $13.1^{+2.6}_{-2.4}\%$, which is in accordance with the predicted runaway Be star fraction when assuming that all of them are binary interaction products. From this they conclude that the kinematic properties of the Be star population are in agreement with population synthesis predictions for post-binary interaction products.

Further observational evidence is given by \citet{Klement2019}, who studied the spectral energy distribution of Be stars searching for possible effects of disruption and truncation invoked by a putative compact companion on the Be star disk. They concluded that many - if not all - Be stars have close companions that influence their outer disks.

Here we propose a novel idea to test the binary hypothesis for Be stars: If the rapid rotation of Be stars originates from previous mass and angular momentum transfer from a binary companion, then there should be an obvious lack of close ($\mathrm{P} \lesssim$\,5\,000\,d) Be\,+\,MS binaries. This is a clear and testable prediction of the binary channel: a lack of Be stars in binary systems with MS companion would strongly support the binary channel as an important formation mechanism of massive Be stars. In contrast, the existence of even one counter example may suffice to show that other mechanisms can also form massive Be stars.

Here, we define close binaries as binaries with current periods $P \lesssim $5\,000\,d. Considering typical radii of red supergiants and typical orbital eccentricities, a maximum initial period of 3\,000\,d is usually assumed for binaries that will interact \citep[e.g.,][]{Sana2012}. However, the orbital period of massive binaries may grow by a factor two or more due to mass loss and mass transfer \citep[e.g.,][]{Vanbeveren1998}. We therefore assume 5\,000\,d to be the limiting period. MS companions around Be stars in longer orbital periods are hence in general not relevant from a binary-evolution perspective.

In the present work, we report on an extensive literature study for a large sample of early-type Be stars in order to investigate the  occurrence of Be stars in close binary systems with a MS companion. Section\,\ref{Sec:sample} describes the sample selection. Section\,\ref{Sec:analysis} presents the analysis of the literature study that we performed. We describe the sample statistics in Section\,\ref{Sec:results}, investigate the effect of detection biases in Section\,\ref{Sec:reasons}, and discuss our results in Section\,\ref{Sec:discussion}. Finally, our conclusions are presented in Section\,\ref{Sec:conclusion}.

\section{Our sample}\label{Sec:sample}
Our aim is to obtain a statistically significant, unbiased sample of classical Be stars. For this, we used the BeSS database\footnote{\textsc{http://basebe.obspm.fr/basebe/}} \citep{Neiner2011}, which is the currently most comprehensive catalogue of classical Be stars. We accessed the continuously updated database on the 18th of August 2019 and selected all classical Be stars with $V\leq12$ and spectral types B1.5 and earlier.

The spectral-type cut enables us to focus our study on the evolution of massive stars ($M_{\rm i} \gtrsim 8\,\mathrm{M}_\odot$). As we show in Appendix~\ref{Sec:Appendix:sample},  in the framework of the binary channel, Be stars with current masses greater than $13\mathrm{M}_{\odot}$ (corresponding roughly to B1.5) should have had a companion  initially more massive than $\mathrm{M}_{i}\gtrsim 8\mathrm{M}_{\odot}$. We stress that lower-mass Be stars may have had massive stars as companions as well, but simple evolutionary arguments ensure this for Be stars with masses exceeding $13\mathrm{M}_{\odot}$.

While the BeSS database is supposed to be complete only down to $V \approx 11\,$mag \citep[][see also Sect.\,\ref{Sec:results}]{Neiner2018}, we include all stars with $V < 12$\,mag. This is mainly to increases our sample size (505 stars instead of 473), and does not alter the results (see Sect.\,\ref{Sec:discussion}).

\section{Analysis of the available literature}\label{Sec:analysis}

\begin{figure} \centering
  \includegraphics[width=0.99\hsize]{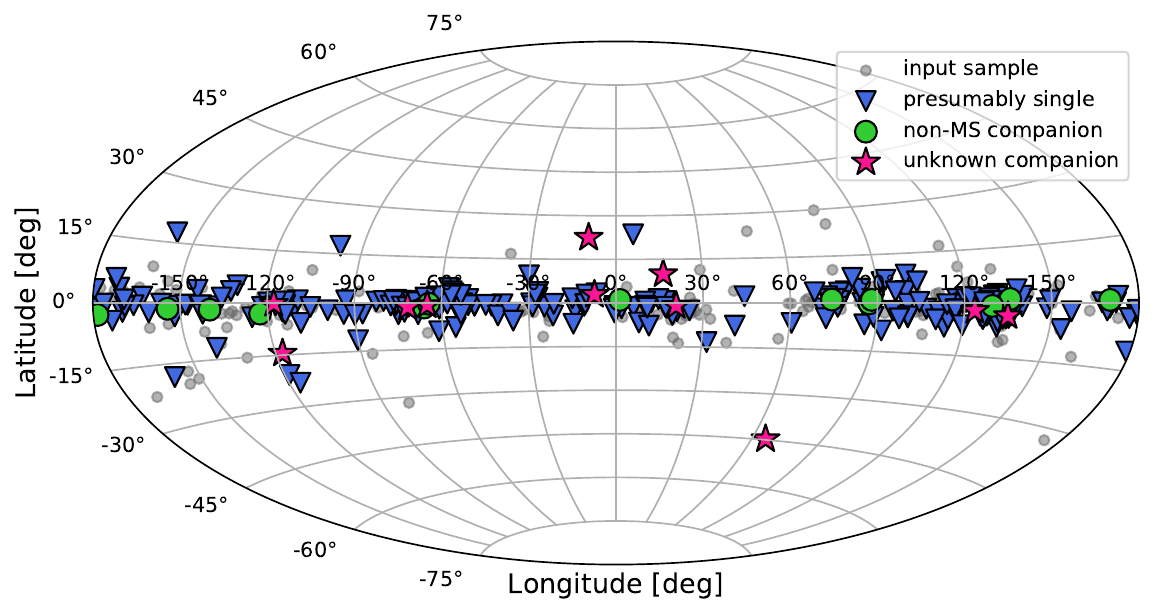}
  \caption{Distribution of our input sample containing 505 stars over the sky (gray dots). The 287 stars in our final sample are marked in color: binaries with known post-MS companions (class i, 13 stars) with green circles, binaries with unknown, uncertain, or debated companions (class ii, 11 stars) with pink stars, and presumably single stars (class iv, 263 stars) with blue triangles. There is no confirmed system of a Be star with MS companion (class iii).}
  \label{Fig:aitoff}
\end{figure}

Our input sample (see Fig.\,\ref{Fig:aitoff}) comprises 505 stars in total for which we perform an in-depth literature research on a star-by-star basis. As BeSS spectral types can be incorrect and often do not provide accurate references, we mainly use SIMBAD \citep{Wenger2007} and the publication history for each star to verify them. Our main sources are \citet{Houk1975}, \citet{Garrison1977}, \citet{Jaschek1982}, \citet{Nesterov1995}, and \citet{Levenhagen2006}. In cases of contradicting spectral types, we chose the one from the most recent traceable reference. In addition we searched the papers published about each object for indicators of binarity \citep[e.g., in][]{Pourbaix2004}, known X-ray sources \citep[e.g., in][]{Naze2018}, known stripped companions \citep[e.g. in][]{Wang2018}, and runaway stars \citep[e.g., in][]{Tetzlaff2011, Peri2012}. We found that results reported in a few studies could not be readily trusted in the context of our study of multiplicity of close Be binaries; these are discussed in Appendix\,\ref{sec:app:Lit}.

Potentially available BeSS spectra were used to provide a quick-look on the spectra (especially to confirm Balmer line emission). For stars in the southern hemisphere, we complemented the BeSS spectra with spectroscopic observations from the ESO archive\footnote{\textsc{http://archive.eso.org/cms.html}}. For stars in the North, we considered archival spectra from the HERMES spectrograph mounted at the KU Leuven Mercator telescope at Roque de los Muchachos Observatory in La Palma, Spain \citep{Raskin2011}, if available.

We removed two groups of stars from the sample. The first group contains stars that are not classical Be stars (see Tab.\,\ref{App:tableNoBe}). These include other emission-line objects such as Herbig Ae/Be stars, B[e] stars, or stars with luminosity class I-II. The group also includes stars without a traceable reference confirming the Be star status, especially if only one historical reference claimed the Be status while other literature from the same period did not report any emission. We also checked archival spectra from BeSS, ESO, and HERMES for possible emission. In this group there are a handful of objects that might, at first glance, appear like Be stars with a close MS companion. However, a careful inspection of the literature and available data reveals that they are actually no classical Be stars. For example \object{CW Cep}, listed as classical Be star of spectral type B1Vve in BeSS was found to be a 2.7d-period B+B binary system with stationary H$\alpha$ emission from circumbinary material \citep{Johnston2019}. We discuss these objects in more detail in Appendix\,\ref{sec:app:reject}.

The second group we removed from the input sample comprises stars without precise spectral classification or stars that turn out to be of spectral type later than B1.5 (see Tab.\,\ref{App:tableLater}). Stars without precise spectral classification (spectral type "Be") are included in the sample because of the setup of the BeSS database search: despite our selection in spectral type, it returns around 150 stars that are classified as "Be" without a more precise spectral subtype. Instead of simply removing these stars from the sample, we searched for spectral classification in the literature, which was available for around 100 of them. Similarly, several stars classified as B1.5 and earlier in BeSS turned out to be of later spectral type in our literature search.
In a similar fashion we might be missing some early-type Be stars: if a star is classified as B2 or later in BeSS it will not be part of our sample, even though it might have a more recent, earlier classification somewhere else in literature. For example, the well-known Be+sdO binary \object{$\phi$ Per} \citep{Poeckert1981} is not included in our sample, since it is classified as B2 in the BeSS archive.

In total there are 50 stars of unknown spectral type, 97 stars that are probably of later spectral type than B1.5, and 71 stars that are unlikely classical Be stars. This leaves us with a final sample of 287 objects. While the vast majority of our objects are of spectral type B0\,V and B1\,V, there are also several O-type stars. We then further classify the 287 confirmed early-type Be stars into one of the following categories:
i) binaries with a confirmed post-MS companion (e.g., NS, sdO);
ii) suspected binaries with companions whose nature is unknown, uncertain, or debated;
iii) Be stars with confirmed MS companions, and iv) presumably single stars, i.e. stars with no indication of binarity, either being single, or with a companion that has eluded detection so far.

The first class of stars (class i) comprises Be binaries for which clear evidence for the presence of a non-MS companion was demonstrated. This includes BeXRBs such as \object{V725 Tau} \citep{Finger1994} or \object{V831 Cas} \citep{Liu2000}. It further includes a handful of stars for which hot evolved companions  were reported based on optical or UV spectroscopy, such as \object{FY CMa} \citep[B0.5\,IV + sdO, ][]{Peters2008} and the WR binary \object{WR\,137} \citep[O9e+WC7, ][]{Lefevre2005}. The properties of this class of stars match the predictions of the binary formation channel, where the original mass donor is now an evolved star or a compact object.

The second class of stars (class ii) comprises all objects for which indications from  spectroscopy, interferometry, or photometry for the presence of a companion exist, but where the nature of the companion is unknown, uncertain, or not agreed upon in the literature. An example is the well-known binary \object{$\gamma$ Cas}, an X-ray bright Be binary with a ${\approx}1\,M_\odot$ companion \citep[see e.g.,][]{Mason1976, Harmanec2000}, which serves as the prototype for the class of $\gamma$ Cas analogues \citep[e.g.,][]{Smith2016}. Despite being one of the brightest stars in the night sky, the nature of the companion remains under debate. Even though the majority of models suggest that the companion is a non-MS companion such as a hot He star or a compact object \citep[e.g.,][]{Postnov2017, Langer2020}, a solar-type companion cannot be fully ruled out. Similar objects include the $\gamma$ Cas analogue \object{$\pi$ Aqr} and the suggested Be+sdO binary \object{HD\,161306} \citep{Koubsky2014}. 
This class also includes the potential Be+MS binary \object{$\delta$ Sco} \citep{Miroshnichenko2013}. All stars in this class are discussed in more detail in Appendix \ref{sec:app:comments}.

The third class (class iii) are Be binaries with confirmed MS companions. Because of the high binary fraction of B stars, and given the high frequency of MS stars in general, B+MS binaries are common (see Sect.\,\ref{Sec:reasons}). One may thus naively expect that this is the most common configuration of Be binaries. As the title of our work implies, this group turned out to contain no stars in our sample. While the presence of an MS companion is compatible with observations of a few Be binaries in our list (most notably $\delta$\,Sco, see Appendix\,\ref{sec:app:unknown}), it was never directly demonstrated (e.g., through spectral disentanglement, SB2 RV curves, or isolated spectral features). For the sake of completeness, we keep this third category.

The fourth class (class iv) contains all Be stars that do not fall in one of the three previous categories, that is, stars for which no indication for a close companion was reported. We refer to these as "presumably single", but stress that this includes binaries with companions that avoided detection, which may or may not dominate this sample.

Naturally, not all targets in our sample were studied equally carefully. Some of the targets are included in systematic RV variability studies, some were the subjects of dedicated studies, while others have been merely classified by several authors. Given the diversity of literature over the past century, and without a systematic spectroscopic multiplicity survey of Be stars, it is difficult to quantify the implied biases due to the large variety of methods and data quality relevant for each target.  We discuss these biases in more detail in Sect.\,\ref{Sec:reasons}.

We give an overview over the entire input sample in Appendix \ref{Sec:Appendix:table}, separated into the final sample of early Be stars (Tab.\,\ref{app:table_sample}), the not-classical-Be stars omitted from our final sample (Tab.\,\ref{App:tableNoBe}), and Be stars of late or unknown spectral type omitted from our sample (Tab.\,\ref{App:tableLater}). The Tables include SIMBAD names, HD numbers, coordinates, V-band magnitudes, spectral types, spectral type references, and individual comments.

\section{Sample statistics}\label{Sec:results}
\begin{figure} \centering
  \includegraphics[width=0.6\hsize]{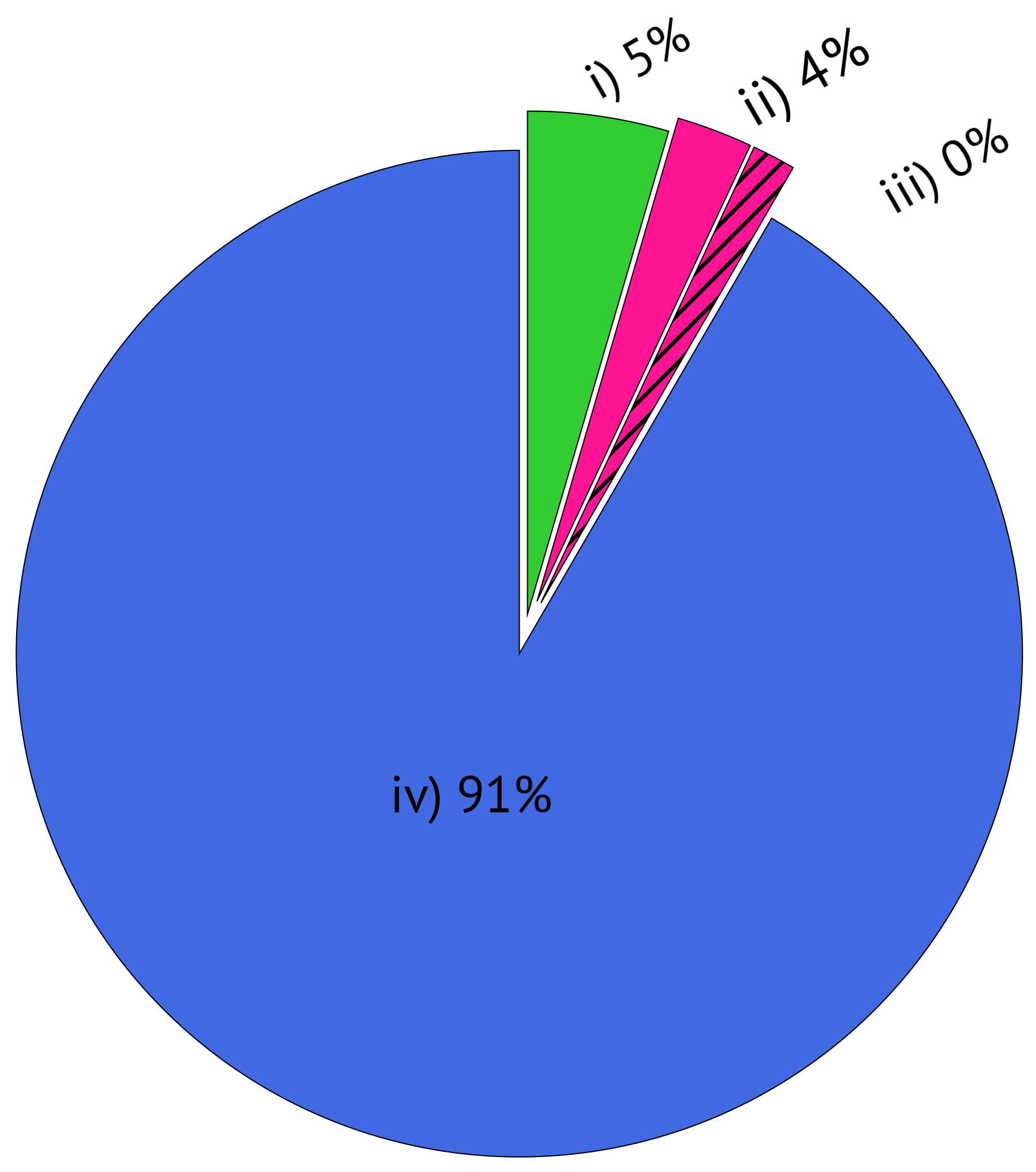}
  \caption{Binary statistics for our final sample of 287 massive Be stars. The vast majority of the stars (263 stars) are presumably single (class iv, blue). Roughly 4.5\% have known non-MS companions (class i, 13 stars, green) while 3.8\% have unknown, uncertain or debated companions (class ii, 11 stars, pink). We subdivide these eleven stars into companion detection methods: seven were studied with spectroscopy (no hatch) while four were detected based on other methods (hatched). There is no report of a Be+MS binary (class iii).}
  \label{Fig:pie}
\end{figure}

In our final sample of 287 Be stars, we find 13 Be stars in a binary system with a non-MS companion (class i) and 11 suspected Be binaries with unknown, uncertain, or debated companions (class ii). Among these eleven stars, seven were first detected spectroscopically and four were detected using other methods. While the presence of an MS companion is  compatible with the observations of a few objects (most notable $\delta$ Sco, see Appendix\,\ref{sec:app:unknown}), we find no Be binaries with confirmed MS companions (class iii). The remaining 263  Be stars are classified as presumably single (class iv), which may be either  truly single or unidentified binaries. The binary statistics of our sample are shown in Fig.\,\ref{Fig:pie}.

\begin{figure} \centering
  \includegraphics[width=0.95\hsize]{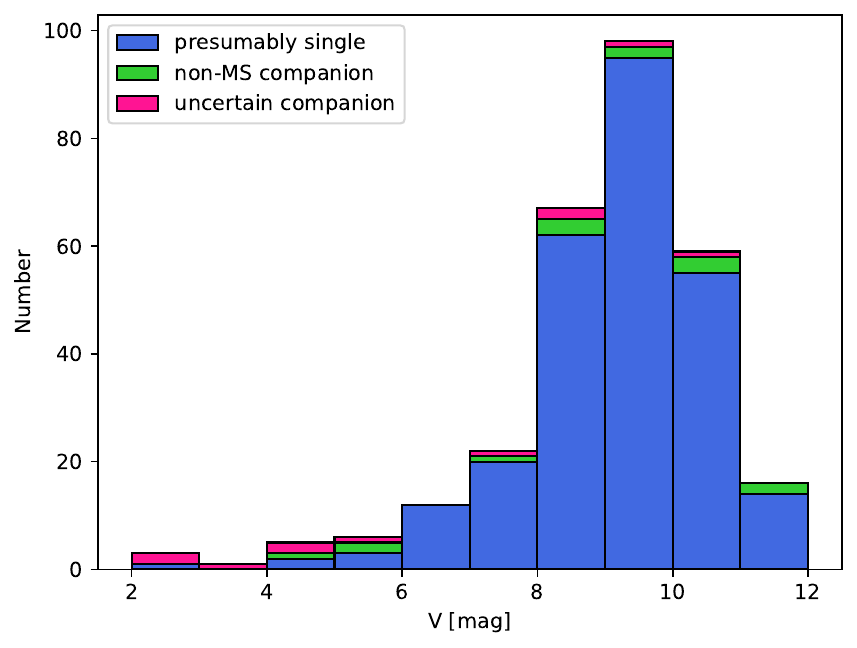}
  \caption{Distribution of V-band magnitudes of the 287 sample stars where class i is shown in green, class ii in pink, and class iv in blue.}
  \label{Fig:maghisto}
\end{figure}

In Fig.\,\ref{Fig:maghisto} we show the distribution of V-band magnitudes in our final sample. The drop in stars at magnitudes fainter than $V=10$ indicates the limit down to which our sample is broadly complete. We therefore test if our binary statistics remain similar when restricting the sample to stars brighter than $V=10$, or $V=11$, following the claimed completeness of the BeSS catalogue. As shown in Fig.\,\ref{Fig:maghisto}, there are known binaries with non-MS companions as well as stars with unknown, uncertain, or debated companions in basically every magnitude bin. Accordingly, our binary statistics do not change when restricting the sample to $V=10$ (class i: 4.7\%, class ii: 4.7\%, class iii: 0\%, class iv: 90.6\%) or $V=11$ (class i: 4.4\%, class ii: 4.0\%, class iii: 0\%, class iv: 91.6\%).

Finding no report of a Be+MS binary is quite remarkable. Typical binary fractions determined for early B-type stars are of the order of $\approx 60\%$ \citep[e.g., ][]{Kobulnicky2014, Dunstall2015}. From an evolutionary perspective, the majority of those binaries should host two MS stars \citep{deMink2014}, suggesting that MS companions to B stars should be quite common. The blatant lack of MS companions to Be stars appears to match with the prediction that the vast majority, if not all, of the Be stars formed through past binary interaction. However, several biases that could reduce the number of reported Be+MS binaries exist. They are discussed in detail in Sect.\,\ref{Sec:reasons}.\newline

\section{Detection biases}\label{Sec:reasons}
Our reported lack of MS companions to Be stars is in agreement with the hypothesis that Be stars are binary interaction products, having gained their rapid rotation due to mass and angular momentum transfer in previous binary interactions.
However, to understand whether the lack of Be+MS binaries in our sample is statistically significant, one should consider possible biases - observational or otherwise - that could contribute to the observed lack of Be+MS binaries.

Given the vast heterogeneity of techniques, data quality, and focus of the many works our study compiles, performing an accurate bias estimate in the context of our study is virtually impossible. A rigorous and robust estimate of the detection biases can only be performed through homogeneous surveys of unbiased and statistically-significant samples of Be stars. Despite this, we try to provide a first order estimate for the overall expected bias against the detection of Be+MS binaries below with the idea to test whether the lack of Be+MS stars in our sample can be explained by a low detection probability or whether it reveals a genuine physical signature. Towards this goal, we adopt a number of conservative hypotheses.

\subsection{Disk truncation and tidal braking}\label{Sec:tidal}

Recent studies by \cite{Klement2019} report on the phenomenon of disk truncation due to the presence of a binary companion. This is supported by the observed lack of short-period (i.e., $P < 30\,\mathrm{d}$) BeXRBs \citep{Raguzova2005}. On the other hand, simulations by \citet{Panoglou2016} indicate that there is no lower limit on the orbital separation for the formation of decretion disks.
In order to be conservative in our estimate we nevertheless assume 
that orbital separations falling short of a certain threshold would disrupt the Be phenomenon.  The typical H$\alpha$ emitting radius of Be disks is of the order of $10-20\,R_\odot$ \citep[e.g.,][]{Rivinius2013}, which corresponds to orbital periods of the order of $10\,$d. 

A reduction of the apparent number of observed Be+B binaries can also be caused by tidal braking \citep{Zahn1977}. In a tight binary system, tides will synchronize the orbital and rotational spins. Rapidly rotating stars are expected to slow down, which may cause a former Be star to not appear as such anymore because of an insufficient rotational speed to sustain its decretion disk. This would reduce the lifetime of the Be+B phase in tight binaries -- or prevent it altogether -- and thus reduce the number of observable Be+B binaries. However, for MS stars with masses of $\approx 15\,M_\odot$, tidal breaking is important on the MS only for tight binaries with periods of the order of $5\,$d \citep{Zahn1977, Hurley2002, Song2013}, and becomes negligible for periods larger than $10\,$d.

Combining the two effects, we therefore assume that no Be+MS binaries with periods $\lesssim 10\,$d would be observed due to disk truncation and tidal braking. Assuming \"Opik's law of period distribution, which is flat on $\log P$ \citep{Opik1924}, this corresponds to a reduction factor of roughly $25$\% in the considered period range. \newline

\subsection{Binary properties of the B star population}
 Present-day, bias-corrected, close-binary fractions larger than 50\% were reported for massive B-type stars in the Milky Way and in the Large Magellanic Cloud \citep[e.g.,][]{Kobulnicky2014, Dunstall2015}. \citet{Kobulnicky2014} report a bias-corrected fraction of 55\% in the period range of $1 < P < 5\,000\,$d and the mass-ratio range of $0.2 < q < 1$. The fraction of reported binaries is fully dominated by MS+MS binaries, since OB+NS binaries will typically have $q \lesssim 0.2$, and OB+BH binaries are very rare \citep[e.g.,][]{Langer2020}. We therefore  assume that the fraction of MS+MS binaries in the aforementioned parameter domain among an unbiased sample of B stars is larger than 50\%.

\subsection{Time sampling and multiplicity studies}
Not all stars in our sample have been equally well studied for multiplicity (see Fig.\,\ref{Fig:maghisto}). The detection of companions relies on various methods, such as spectroscopy, interferometry, photometry, and imaging. Of those, spectroscopy has been most predominantly used for binary detection, especially when focusing on the period range of interest, i.e. $P<5000$\,d. Only a few Be stars were studied interferometrically \citep[e.g., $\tau$\,Sco,][]{Tango2009}, and imaging studies, as argued in Sect.\,\ref{Sec:intro}, generally do not probe separations that are relevant in the framework of binary interaction. We therefore focus our bias-estimate discussion on spectroscopy.

In principle, if an MS companion is bright enough, its contribution may be seen in the spectrum without requiring multi-epoch spectroscopy. However, it is difficult to estimate when such a companion would be visible: this depends not only on the data quality, spectral coverage, and spectral appearance of both components, but also on the researcher and their goals in the respective studies. We therefore conservatively assume that Be binaries with MS companions would only be identified through multi-epoch spectroscopy. For this reason, we now estimate the fraction of Be stars in our sample for which multi-epoch spectroscopy has been performed.

Again, this is difficult to do on the basis of our literature studies, primarily since systematic multi-epoch publications of Be stars are limited. A powerful bias that needs to be considered is the fact that researchers in the field have rarely published non-detections. For example, studies based on data acquired with the Ond\u{r}ejov observatory report on quite a few unique systems, for example on Be binaries with sdO companions, systems with unidentified companions, or systems that do not contain Be stars after all \citep[see e.g.,][]{Harmanec2002, Saad2004, Linnell2006, Koubsky2010, Koubsky2019}. However, no studies on non-detections were published. Our knowledge of multiplicity among Be stars generally relies on individual, star-by-star analyses.

Here, we assume that the more frequently a Be star has been observed by the community, the more likely it is that the star was actually studied based on multi-epoch spectroscopy. The BeSS catalogue lists the number of spectra available per object, where the sources of the spectra vary. We assume that the number of individual observations on the BeSS archive can be used as a proxy for whether or not the star was studied via multi-epoch spectroscopy. For our final sample of 287 Be stars, 49 stars have more than 10 unique spectroscopic observations. In this counting we take into account the fact that the BeSS archive sometimes lists different echelle orders as individual spectra, which would lead to a double-count. On this basis, we estimate that approximately 49/287 ($\approx$ 17\%) of our sample were studied by means of multi-epoch spectroscopy, and we assume that at least 10 epochs of spectra were secured. This does not imply that the BeSS spectra were actually analysed for multiplicity. Rather, we use this as a proxy to estimate  the fraction of Be stars in the sample that was well-studied by the community.

\subsection{Observational biases}\label{sec:obsbias}

\begin{figure}
  \includegraphics[width=.54\textwidth]{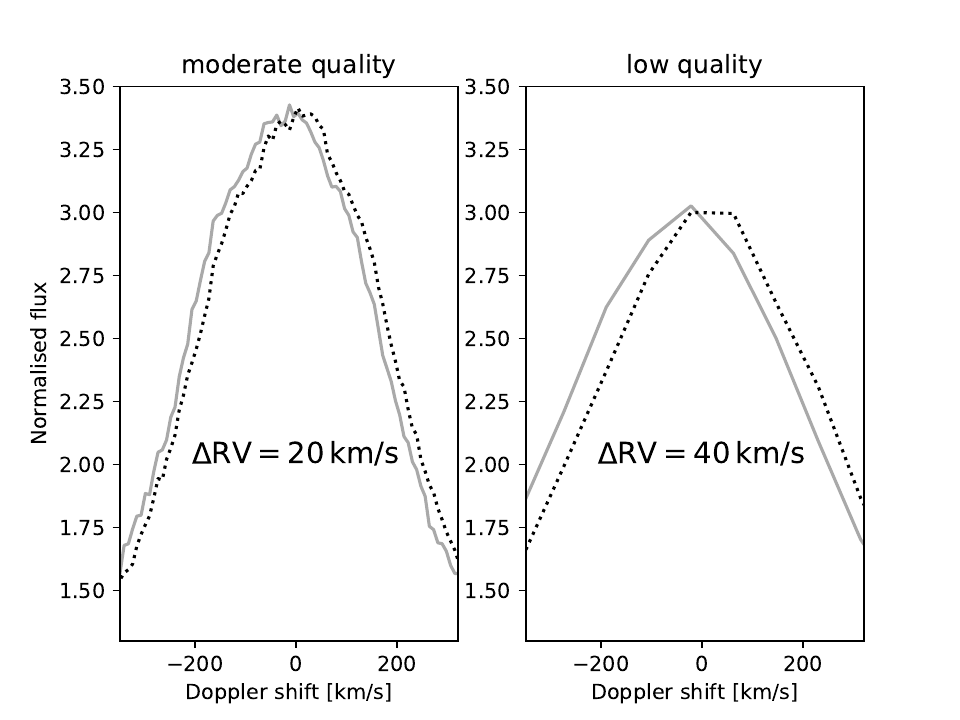}
  \caption{Simulated Doppler-shifts of the H$\alpha$ line profiles of the Be star $\gamma$~Cas for the case of moderate-quality  (left, $R =10\,000$, S/N=50, $\Delta {\rm RV} = 20\,$\kms) and low-quality (right, $R=1000$, S/N=30, $\Delta {\rm RV} = 40\,$\kms) spectra. The Figure illustrates that such shifts can be easily identified in the spectrum, without relying on sophisticated RV measurement techniques.}
  \label{Fig:SpecSim}
\end{figure}

We now estimate the fraction of SB1 or SB2 detections in a sample of Be stars studied via multi-epoch spectroscopy. Binaries are flagged on the basis of significant RV changes. Again, the threshold depends on the data quality and spectral features considered. For simplicity, we consider the H$\alpha$ line, 
which provides the highest signal-to-noise due to the strong emission,
and two limiting cases:  moderate-quality data ($R = 10\,000$, S/N=50) and low-quality data ($R=1\,000$, S/N=30). In Fig.\,\ref{Fig:SpecSim}, we show observed H$\alpha$ profiles of the Be star $\gamma$\,Cas, degraded to the respective data quality considered here, and artificially shifted in RV. Figure\,\ref{Fig:SpecSim} illustrates that a RV shift of $20\,$\kms~(40\,\kms) would be readily seen using moderate-quality (low-quality) spectra of H$\alpha$, without relying on sophisticated RV measurement techniques. For the subsequent estimates, we assume that objects in which the maximum RV shift, Max($\Delta$ RV), is larger than these respective values, would be flagged as a binaries.

We note that this shift in RV is a significantly different signature than what is expected from variability in the disk, which would not shift the entire line but rather affect its strength and shape. Pulsations can also cause line profile variability that might mimic RV shifts, especially in He\,\textsc{i} absorption lines. The RV shifts are, but for a few cases \citep[see e.g.,][]{Bolton1982, Baade1982}, typically of the order of a few km/s \citep{Aerts2009} and thus below the RV thresholds of 20 and 40 km/s that we consider here. Mistaking the RV signature of pulsations for an orbital motion in a binary would, however, lead to false-positive detections rather than an observed lack of Be binaries. Our thresholds are also in line with typical binary detection thresholds of $\lesssim$20\,\kms~used in the literature \citep{Sana2011, Sana2012, Sana2013, Dunstall2015}.

We next convert the Max($\Delta$ RV) threshold into corresponding RV amplitudes of the Be star. That is, we want to understand to which RV amplitudes $K$ typical multi-epoch studies would be sensitive to. In principle, the larger the number of epochs, the closer Max($\Delta$ RV) would approach the full RV amplitude $2\,K$. To estimate Max($\Delta$ RV) as a function of the number of epochs $N$, we perform Monte Carlo simulations by randomly sampling $N$ RV measurements of a binary orbit. For $N=10$, one obtains a mean maximum shift of Max$(\Delta {\rm RV}) \approx 1.7\,K$ for eccentricities of $e=0.5$. For circular orbits, the ratio becomes larger ($\approx 1.85\,K$), while it becomes roughly $1.5\,K$ for $e=0.8$ and approaches zero for arbitrarily large eccentricities. For simplicity, we assume an orbit with $e=0.5$, and hence adopt Max($\Delta$ RV) = $1.7\,K$. By replacing the left hand side of the equation of our estimates (20 and 40 \kms~for moderate- and low-quality spectra, respectively), we find that the moderate-quality data would be sensitive to RV amplitudes $K \gtrsim 12\,$\kms, while low-quality data would be sensitive to $K \gtrsim 24$\,\kms , which is consistent with results reported in past spectroscopic surveys \citep[e.g.,][]{Abt1978}. 

\begin{figure} 
  \includegraphics[width=.54\textwidth]{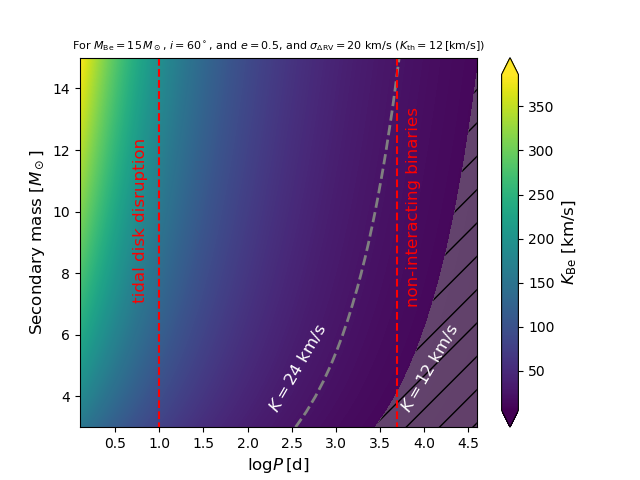}
  \caption{Sensitivity map of multi-epoch spectroscopic surveys. Color depicts the RV amplitude of a $15\,M_\odot$ Be star in a binary with $i=60^\circ$ and $e=0.5$ as a function of the orbital period and secondary mass. Over-plotted are contours for $K_{\rm Be} = 12,24\,$\kms~(see text). }
  \label{Fig:TomerPlot}
\end{figure}

Our sample is composed almost entirely of B0\,V and B1\,V stars in roughly equal numbers. Let us therefore assume  that all stars in our sample are of spectral type B0.5\,V, corresponding roughly to  $M\approx 15\,M_\odot$. For simplicity, we again assume an eccentricity of $e=0.5$. Figure\,\ref{Fig:TomerPlot} shows the RV amplitude of the Be component in such a binary with an inclination of $i = 60^\circ$ as a function of the orbital period and secondary mass.
In the plot, we mark the periods below which disks are not expected to exist due to tidal disruption (see Sect. \ref{Sec:tidal}), and beyond which binary interaction is assumed to be negligible ($P \gtrsim 5\,000$ d). Based on the estimates above, we plot contours of $K_{\rm Be} =12$ and $24$\,\kms. The y-axis covers mass ratios between 0.2 and 1, which overlaps with the parameter range adopted by \citet{Kobulnicky2014}. Since the probability distributions of the mass-ratio and of $\log P$ are found to be approximately flat \citep[e.g.,][]{Sana2012, Kobulnicky2014}, the area on the diagram is proportional to the number of objects that are expected to populate the corresponding parameter domain. Hence, Fig.\,\ref{Fig:TomerPlot} offers a simple way of estimating the number of SB1 or SB2 systems that would be detected. For example, for $i=60^\circ$, moderate-quality data would be sensitive to almost the entire relevant parameter domain of $1<P<5\,000$\,d and $0.2<q<1$ ($\approx 99.8\%$), while low-quality data are sensitive to roughly 90\% of this  parameter domain. For lower inclinations, the sensitivity domains becomes smaller. For example, for $i=30^\circ$, low-quality data are sensitive to $\approx 70$\% of the domain. Weighing these sensitivity fractions with the probability density of randomly aligned inclinations ($p(i) = \sin i$), one obtains an overall sensitivity of 84\% for low-quality data and 96\% for moderate-quality data.

\section{Discussion}\label{Sec:discussion}

Considering all effects together, we can estimate the expected number of Be+MS stars with $0.2<q<1$ and $1<P<5\,000$\,d that would be detectable as SB1 or SB2 binaries in our sample. We obtain a total reduction factor of (disk truncation factor) $\times$ (fraction of binaries in parameter domain) $\times$ (fraction of binaries studied through multi-epoch spectroscopy) $\times$ (sensitivity of multi-epoch spectroscopy) $\gtrsim$ 0.75 $\times$ 0.5 $\times$ 0.17 $\times$ 0.84 = 5.4\%. The number grows to 6.1\% when assuming moderate-quality data (sensitivity of 0.96, see Sect.\,\ref{sec:obsbias}). That is, if Be stars would follow standard B-star statistics, we estimate that 5-6\% of our sample would have been reported as SB1 or SB2 binaries with periods and mass ratios in the considered ranges, which corresponds to 15-18 stars in our sample of 287 Be stars.

As reported in Sect.\,\ref{Sec:results}, we find no unambiguous report of a Be+MS binaries. We identify a total of seven SB1 or SB2 Be binaries with uncertain companions: $\delta$\,Sco, HD\,93683, EM*\,MWC\,711, and $o$\,Pup, HD\,1613606, $\gamma$\,Cas, and $\pi$\,Aqr. However, the latter four targets have reported mass ratios smaller than 0.2 (see Sect.\,\ref{sec:app:comments}) and should therefore be excluded from this comparison because they are outside of the considered parameter space.  Hence, in our total sample of 287 objects, merely three objects are potential Be+MS binaries in the considered parameter domain, in contrast to the expected value of 15-18. Assuming binomial statistics, the likelihood of observing three SB1 or SB2 binaries or less for an expected value of 15 (our lower limit) is very small (p $\sim 10^{-4}$). Even if we consider the low mass-ratio systems, the probability of detecting seven binaries or less by chance would be $\approx$0.02.

\begin{figure} \centering
  \includegraphics[width=0.99\hsize]{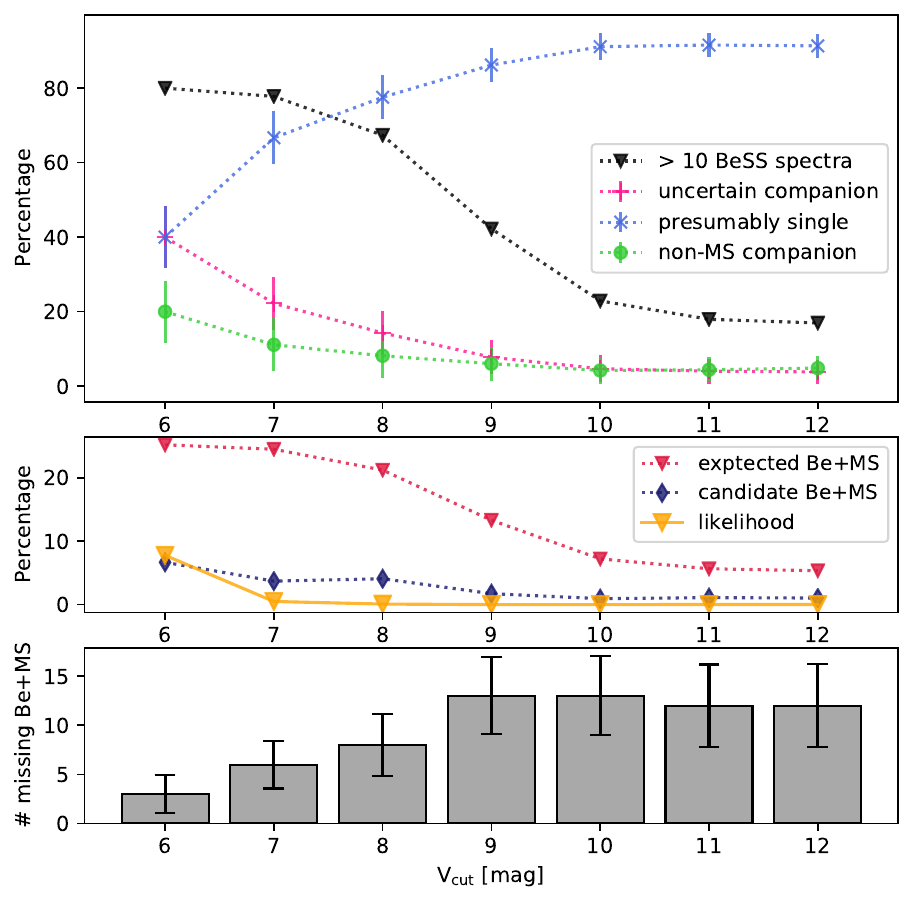}
  \caption{Top panel: Dependence of the different fractions and their binomial errors reported in Sect.\,\ref{Sec:results} as a function of the magnitude cut. Also indicated is the fraction of stars in the sample that have more than 10 individual BeSS spectra available. 
  Middle panel: Comparison between the expected fraction of Be+MS systems and the observed fraction of possible Be+MS candidates. In addition we show the corresponding binomial likelihoods of our findings being the result of chance (see text). Bottom panel: number of missing Be+MS systems as a function of the magnitude cut.}
  \label{Fig:fractions}
\end{figure}

In Fig.\,\ref{Fig:fractions} we further test the robustness of our results by investigating whether the lack of detected Be+MS systems depends on the magnitude cut of our sample. We thus consider sub-samples with different cuts in the V band magnitude and compute the above-mentioned fractions and likelihood assuming low-quality data (i.e. a sensitivity factor of 0.84).
Fig.\,\ref{Fig:fractions} indicates that the fractions considered do change for different input samples. This is, however, taken into account in the estimate of the detection biases. 
As Fig.\,\ref{Fig:fractions} indicates, the lack of reported Be+MS systems is significant regardless of the magnitude cut of our sample.

The simplistic but conservative bias estimate therefore strongly suggests that this result is unlikely to be due to chance, and is likely rooted in a distinct evolutionary path that massive Be stars tend to follow. Taken at face value, to reduce the expected number of $\approx$ 15 Be+MS stars in the considered parameter domain to one that would be compatible with literature (three or smaller), binary interactions would need to be responsible for the formation of at least 75\% of the early-type Be stars. 

We acknowledge the fact that obtaining robust bias estimates for such a heterogeneous data set is virtually impossible. Our estimates above should only be considered as a rough order-of-magnitude estimate, and especially the fraction of Be stars forming via the binary channel should be taken with a grain of salt. Our intention in this study is to report on the blatant lack of Be+MS binaries in the literature: it is beyond the scope of our work to pursue possible biases any further. To obtain solid bias estimates systematic, homogeneous surveys of statistically significant samples of Be binaries are required.

\section{Conclusion and future work}\label{Sec:conclusion}
In this study, we investigate the hypothesis whether the majority, if not all, of the massive  Be stars ($M_{\rm Be} \gtrsim 13\,M_\odot$) have formed through binary mass-transfer. We argue that, if true, there should be a clear lack of close ($P \lesssim 5\,000$\,d) Be + MS binaries. We performed an extensive literature study of a sample of 505 Be stars with $V\leq12\,\mathrm{mag}$ listed in the BeSS catalogue \citep{Neiner2011}. After removing non-classical Be stars and Be stars of unknown spectral types or spectral types later than B1.5, we established a final sample of 287 early-type Be stars. 

Among this sample of 287 Be stars, 13 are confirmed Be binaries with non-MS companions, 11 are binaries with unknown, uncertain or debated companions, and the remaining 263 Be stars are presumably single (including both truly single Be stars or binaries in which the companion avoided detection). Importantly, none of the objects in our sample were reported as confirmed Be+MS binary.

Despite the inhomogeneity of our sample and the resulting difficulties to perform a careful bias estimate, we demonstrated that this blatant lack of Be+MS binaries is unlikely to be a consequence of observational or statistical biases. We therefore suggest that it is a genuine consequence of stellar evolution. The fact that a handful of Be+sdO systems are known despite the large bias against their discovery makes this even more evident.

Comparing our results to the expected number of detectable Be+MS binaries after accounting for biases might suggest that a vast majority of the early-type Be stars formed through binary mass-transfer. This, on the other hand, would imply that a majority of the early-type Be stars should be in binaries with helium stars or compact objects, or else they are single stars disrupted in a supernova event.

Future homogeneous multi-epoch surveys should be dedicated to test this prediction and to establish the distribution of binary companions of Be stars. Large-scale spectroscopic surveys such as APOGEE or LAMOST can profoundly improve this situation by providing us with a homogeneous database of high-quality spectra.

\begin{acknowledgements}

The authors would like to thank the referee, Dr. Dietrich Baade, for his comments that greatly improved our work. Furthermore, we would like to thank Norbert Langer, Lidia Oskinova, and Dominic Bowman for very helpful discussions.
The authors acknowledge support from the FWO Odysseus program under project G0F8H6N and from the European Research Council (ERC) under the European Union’s DLV-772225-MULTIPLES Horizon 2020 research and innovation programme. This research has made use of the SIMBAD database, operated at CDS, Strasbourg, France, the BeSS database, operated at LESIA, Observatoire de Meudon, France, and of NASA's Astrophysics Data System Bibliographic Services.
\end{acknowledgements}

\bibliographystyle{aa}
\bibliography{papers}

\begin{appendix}

\section{Selection limits on spectral subtype}\label{Sec:Appendix:sample}
In the present study, we limit our sample selection to stars with present-day spectral type B1.5 and earlier. This criterion is based on the desire to restrict our study to massive primaries, i.e. primaries with initial masses $>8\mathrm{M}_{\odot}$. The correspondence between present-day spectral type and initial mass follows from the following simple evolutionary considerations.\newline
The initial mass of the primary $M_{ini,1}$ is by definition larger than or equal to the initial mass of the secondary $M_{ini,2}$:
\begin{equation}
  M_{\rm ini,1} \geq M_{\rm ini,2}\,.
\end{equation}
Let us define $\epsilon$ as the mass transfer efficiency, $f \cdot M_\text{ini,1}$ the mass of the primary's core (which is retained during mass transfer), and $(1-f) \cdot M_{\rm ini,1}$ the mass of the primary's envelope (which is removed during mass transfer). The mass of the secondary accreting a fraction $\epsilon$ of the transferred mass, i.e. the mass of the Be star we observe, is then:
\begin{equation}
  M_{Be} = M_{\rm ini,2} + \epsilon \, (1-f) M_{\rm ini,1} \,.
\end{equation}
From this follows
\begin{equation}
  M_{\rm ini,2} = M_{Be} - \epsilon \, (1-f) M_{\rm ini,1}
\end{equation}
and using Eq. A1:
\begin{equation}
  M_{\rm ini,1} \geq M_{Be} - \epsilon \, (1-f) M_{\rm ini,1}\,.
\end{equation}
Given that the mass transfer efficiency $\epsilon \leq 1$, the initial mass of the primary is
\begin{equation}
  M_{\rm ini,1} \geq \frac{M_{Be}}{1+\epsilon\, (1-f)} \geq \frac{M_{Be}}{2-f} \,.
\end{equation}
Hence, to guarantee that $M_{\rm ini, 1} \ge 8\,M_\odot$, we require:
\begin{equation}
  M_{\rm Be} \geq 8\cdot(2-f)\quad [M_{\odot}] \,.
\end{equation}

Assuming a typical value for the core mass fraction of \mbox{$f\approx 0.33$} \citep[e.g.,][]{Limongi2018}, we  find that a massive primary is ensured for a minimum current mass of the Be star of $\approx 13\mathrm{M}_{\odot}$. This in turn roughly corresponds to a spectral type of B1.5 \citep{Silaj2014}. We stress that it is  possible for lower-mass Be stars to have had massive stars as companions. However, in the framework of the binary channel, Be stars more massive than approximately $13\,M_\odot$ must have had massive companions.

\section{Comments on the literature and individual stars}\label{sec:app:comments}
\subsection{Literature selection}\label{app:literature}
\label{sec:app:Lit}

As described in Sect.\,\ref{Sec:intro}, we performed an extensive literature research encompassing many different techniques, instruments, and methodologies. Doing so, we tried to maintain an unbiased and inclusive approach. However, a few studies turned out to be unreliable for studying the multiplicity of close ($P \lesssim 5\,000\,$d) Be binaries, and those are outlined below.

\citet{Oudmaijer2010} performed an imaging survey of Galactic B and Be stars and reported on the presence of visual companions. However, their study was sensitive to angular separations of at least 0.1'', corresponding to minimum orbital separations of at least 20\,au, typically of the order of \mbox{a few hundred au}. At such large separations the components are irrelevant from the perspective of binary evolution. Therefore, binaries or multiples reported by \citet{Oudmaijer2010} are not considered as close binaries in our study.

\citet{Chini2012} performed a multi-epoch spectroscopic study of massive stars and classified them into "constant," "SB1," and "SB2," depending on the spectral variability and line morphology. We found that several targets in our sample were classified as "SB2" despite not having been reported as such in other studies. A blatant example is the bright star \object{$\zeta$\,Oph}, which is a runaway that was the subject of multiple dedicated spectroscopic campaigns \citep[e.g.,][]{Reid1993,Oskinova2001}. \citet{Chini2012} considered all objects as SB2s that show spectral line deformations. As the majority of Be stars are known to pulsate and exhibit spectral variability unrelated to orbital motion, this condition cannot be used to classify Be stars as binaries. While we document binary classifications by \citet{Chini2012} in Tables \ref{app:table_sample}, \ref{App:tableNoBe}, and \ref{App:tableLater} (19 detections overall), we do not consider these targets as suspected binaries.

\citet{Kervella2019} recently provided a comprehensive list of binary candidates based on a combined analysis of Gaia and Hipparcos coordinates and proper motions. Their study focused on low-mass nearby stars. However, Hipparcos measurements of massive stars are known to have calibration issues \citep[e.g.,][]{Schroder2004, Hummel2013, Shenar2015}. Moreover, the majority of targets reported as multiple by \citet{Kervella2019} are very close to the threshold of not being considered significant detections. We therefore do not consider targets flagged by \citet{Kervella2019} as suspected binaries. However, we do mentioned positive detections in Tables \ref{app:table_sample}, \ref{App:tableNoBe}, and \ref{App:tableLater}.

\subsection{Binaries with unknown, uncertain, or debated companions}
\label{sec:app:unknown}
Eleven stars in our sample appear to be in a binary systems with an unknown, uncertain, or debated companion (class ii). They are detected by different observing methods and some of them are significantly better studied than others. In the following we briefly comment on all of them.

\paragraph{$\gamma$ Cas} is one of the brightest stars in the night sky and the first Be star ever discovered. It was found to emit hard, moderately strong, and thermal X-ray radiation on the basis of which it serves as the prototype for the class of $\gamma$ Cas analogues \citep[e.g.,][]{Mason1976, Naze2018}. \cite{Harmanec2000} measured a long-term RV curve indicative of $\gamma$ Cas being in a binary system with low-mass companion on a 203 days period. The nature of this companion is, however, not well constrained. The companion was speculated to be a white dwarf \citep{Haberl1995}, a NS \citep[e.g.,][]{Postnov2017}, or a He star \citep[e.g.,][]{Langer2020}. 
A recent study by \citet{Borre2020} studied the long-term variations of the H$\alpha$ profile and found indications that a spiral structure in the circumstellar decretion disk is in phase with the companion.
However, it cannot be fully ruled out that the companion is a MS star. Therefore, to remain conservative, we classify \object{$\gamma$ Cas} as a binary with still debated companion (group ii) rather than a non-MS companion, group (i).

\paragraph{$\pi$ Aqr,} a B1e star, is a $\gamma$ Cas analogue \citep{Naze2017} based on its X-ray emission. \cite{Bjorkman2002} report on a companion on a 84.1 days orbital period with a minimum mass of ${\approx}2\,M_\odot$. They find a trailing H$\alpha$ emission feature which they interpret to be indicative of an A- or F-type MS companion based on evolutionary arguments. However, the companion may equally well be a relatively massive He star \citep[e.g.,][]{Langer2020}. \cite{Zharikov2013} confirm the mass of the secondary to be around ${\approx}2\,M_\odot$ which they interpret as too high for an sdO or sdB star. \cite{Klement2019} detect no SED turndown which would be expected for a close companion.

\paragraph{$\delta$ Sco} is classified as B0.2\,IV \citep[e.g.,][]{Chini2012} and was confirmed to be a highly eccentric ($e = 0.94$) binary with an orbital period of 10.8\,yr through spectroscopic \citep[e.g.,][]{Miroshnichenko2001} and interferometric \citep[e.g.,][]{Tango2009, Meilland2011} studies. Through interferometry, the companion was found to be roughly five times fainter than the Be primary in the visual \citep{Tango2009}. Assuming the companion is an MS star, \citet{Tango2009} estimated masses of $M_{\rm Be} = 15\pm7\,M_\odot$ and $M_{\rm comp} = 8.0\pm3.6\,M_\odot$. \citet{Meilland2011} also suggested that the companion may be a B-type star with a spectral class ranging between B2\,V and B4\,V. \cite{Miroshnichenko2013} confirmed the mass estimates and present a radial velocity curve for the primary. While they interpret variability in the wing of the He\,I line at 4471$\AA$ as a signature of the secondary component and find it to be consistent with an early B-type star, they cannot unambiguously characterize it. Due to long-term deviations in the measured RVs, they propose that the system may be a runaway triple system with a third, yet undetected companion. $\delta$\,Sco should be subject to additional high S/N spectroscopic campaigns to confirm or reject the presence of a MS companion. If the companion can be confirmed to be a MS star, \object{$\delta$ Sco} would become the smoking gun for the formation of Be stars from the single-star channels.

\paragraph{V1075 Sco} is the earliest star in our sample, classified as O7.5\,V((f)) by \citet{Sota2014}.  \citet{Negueruela2004a} were inconclusive regarding whether or not \object{V1075 Sco} should be considered an Oe star, based on various line morphologies. Through an interferometric survey, \citet{Sana2014} discovered a companion at a separation of about 25\,mas with a magnitude difference of roughly 0.3\,mag. The separation corresponds to orbital periods of the order of 5\,000\,d, which is close to our (conservative) upper limit of post-interaction binaries. Given its brightness, the secondary may well be a MS star of a similar spectral type. \citet{Chini2012} classified the system as "SB2". However, as we discuss in Sect.\,\ref{sec:app:Lit}, these authors generally consider variable stars as SB2. Moreover, they do not specify the nature of the secondary they claim to see in the spectrum. We therefore do not consider \object{V1075 Sco} as a confirmed Be+MS binary. Like $\delta$ Sco, \object{V1075 Sco} should be subject to further monitoring due to its potential importance in demonstrating the possible formation of Be star through single-star channels.

\paragraph{HD 93683} was roughly classified as OBe in \cite{Stephenson1971} and as B0/1(V)ne in \cite{Houk1975}. The most recent spectral classification,  O9V+B0V \citep{Alexander2016}, implies a binary configuration. Interestingly, these authors report that there is no evidence of emission lines in the spectrum. However, archival spectra from FEROS at the 2.2m telescope in La Silla Observatory, Chile, show strong emission lines not only in the Balmer lines but also in He I and Fe II lines. \cite{Otero2006} find the system to be an eclipsing binary with a period of 18 days. It is associated with a stellar bow shock visible in the infrared \citep{Sexton2015, Kobulnicky2016}. Recent observations from the Shenton Park Observatory (SPO) taken by Paul Luckas using a 0.35m Ritchey-Chr\'etien telescope equipped with a Shelyak Lhires iii spectrograph covering the spectral range around H$\alpha$ ($6500-6610\AA$) show that the H$\alpha$ line, which is significantly shifted in RVs in 2015 and 2016, does not follow the movement of any of the two stars in the system. We therefore propose that there is a third component in the system hosting a disk in which the H$\alpha$ emission is formed. It traces a different orbit than the 18-day eclipsing binary, with an estimated period of the order of 400 days. While this component may be a Be star, we cannot identify photospheric spectral features that can be unambiguously attributed to such third B-type star. Future long-term monitoring of the system is required to investigate the nature of the putative third component and the origin of the H$\alpha$ emission.

\paragraph{$\kappa$ CMa} was classified as B1.5Ve by \citep{Levenhagen2006}. \citet{Klement2019} detect an SED turndown indicative of of the presence of a close companion. The nature of the companion was, however, not confirmed. $\kappa$ CMa was also studied by \citep{Wang2018} who only find a signature of the primary. \cite{Eggleton2008} find no indication for the star being a binary. Future observations need to clarify the possible presence and nature of a close companion to $\kappa$ CMa.

\paragraph{V916 Cen} was reported to be a Be star by several authors \citep[][ and references therein]{Skiff2014}. The two archival FEROS spectra show no emission, and no additional BeSS spectra are available. The object was classified as eclipsing binary as well as $\beta$ Cep and $\lambda$ Eri variable by \cite{Pigulski2008}. They report an orbital period of 1.46 days while the shape and depth of the eclipses imply a companion of similar temperature and size. \cite{Mayer2016} argue, based on the lack of RV variations in their spectra, that the lightcurve could also show typical Be star variability rather than show the signature of an eclipsing binary. The two archival FEROS spectra spectra are taken close to phase 0 so that no additional information about the binary status can be drawn from them. It is possible that the deblending of Balmer lines in spectra taken at quadrature might have been mistaken as line-infilling characteristic of the Be phenomenon.
 
\paragraph{V494 Sct} was classified as eclipsing binary based on an ASAS lightcurve by \cite{Williams2011}. They report an orbital period of 4.95 days and an inclination of 81.3$^{\circ}$. The system was classified as "detached Main-Sequence system" by \cite{Avvakumova2013}. Due to the lack of spectra we cannot investigate the nature of the companion. However, it is difficult to conceive how a decretion disk should form around one of the companions in a detached system. It is possible that the star was classified as Be star, for example, due to the variable nature of the system or a possible circumbinary disc, but we keep it in the sample of stars with unknown companions until further studies clarified the nature of the system.

\paragraph{EM* MWC 711} is a member of the h and $\chi$ Persei double-cluster. The most recent spectral type as well as a confirmation of the Be nature were reported by \cite{Mathew2011}. \cite{Strom2005} list the star as candidate binary because the star's RV amplitude differs from the cluster mean velocity by $\Delta \mathrm{v} = -44\,\mathrm{km/s}$. As the authors do not report on double lines indicative of an SB2 the star is classified as SB1. However, it could equally be a runaway.

\paragraph{$o$ Pup} is an early type star of spectral type B1 IV:nne. \cite{Koubsky2012} measured RVs from Balmer and He\,{\sc i} emission lines and found that they move in anti-phase with a period of 28.9 days. Based on the inferred mass ratio and spectral properties, they suggest that $o$\,Pup is a Be+sdO binary. In contrast, the companion in the Be binary $\pi$\,Aqr (see above), which shows similar a spectral behaviour and mass ratio, was interpreted as a Be+MS binary. We therefore assign $o$\,Pup, like $\pi$\,Aqr, to class ii (binaries with uncertain or debated companions).

\paragraph{HD 161306} was classified as B0:ne. Much like $o$\,Pup (see above), 
\cite{Koubsky2014} report a possible sdO companion orbiting the Be star at a period of about 100\,d. Like $o$\,Pup and $\pi$\,Aqr, we therefore assign this object to class ii (binaries with uncertain or debated companions).

\subsection{Stars rejected as Be+MS binaries}\label{sec:app:reject}

In the following, we discuss objects that appear like Be+MS binaries at first glance. Taking a closer look, however, all of these stars turned out to be very likely false classifications of classical Be stars, primarily due to the confusion of a circumstellar disk with a circumbinary disk, or due to line-deblending in spectra of close binary systems taken at quadrature phases. They are thus removed from the final sample.

\paragraph{CW\,Cep} is a detached, double-lined, eclipsing binary system with a period of 2.7 days. Both components are of spectral type B0.5V and have similar masses \citep[$M_1 = 11.82 - 13.49 M_{\odot}$ and $M_2 = 11.09 - 12.05\,M_{\odot}$,][]{Petrie1947, Popper1974, Clausen1991, Han2002a}. \citet{Johnston2019} use high-resolution optical spectroscopy to determine orbital as well as atmospheric parameters for both components. They report stationary H$\alpha$ emission, i.e. the emission line is not shifted according to the orbital motion of each of the two stars. From this, the authors conclude that the emission does not originate from a decretion disk around one of the two stars but could originate from circumbinary material, i.e. in a circumbinary disk or envelope. Hence, the binary cannot be claimed to consist of a Be component.

\paragraph{HD 17505} is a triple (or quadruple) system containing at least three O-type stars \citep{Sota2014} that was very recently reported to show no emission \citep{Raucq2018}. The last time it was classified as emission line object was by \citet{Hardorp1959} and the emission was not confirmed in any other study. It is very likely that the multiple nature of the system was misinterpreted as infilling of the H$\alpha$ line, and confused with the Be phenomenon.

\paragraph{SX Aur} is a contact binary system of a B2V and a B4V star \citep{Ozturk2014}. Given its spectral type, the object is not considered in our final sample of early-type Be stars. Moreover, it is not conceivable how a disk should exist around one of the components in a contact system. It is therefore much more likely that the emission originated in a cirumbinary disk, or that the deblending of the Balmer lines at quadrature was confused with the Be phenomenon.

\paragraph{HD217061} is a spectroscopic binary system with a 2.6 d period. While the primary is a B1V star, a classification of the second star was not possible so far. The emission was only reported by \cite{Hardorp1964} and not confirmed since. \cite{Pourbaix2004} argue that the emission was likely confused with the SB2 nature of the object. We therefore do not consider this object a classical Be star, much like \object{CW\,Cep}.

\paragraph{RY Sct} is a rare post-mass transfer system reported by \cite{Grundstrom2007} who argue that the system might evolve into a Be + WR binary. Currently, however, the B star is hidden by the brighter primary, and \citet{Grundstrom2007} suggest that no component in the system can be assigned the spectral class Be. We therefore omit this object from our final sample of Be stars.

\paragraph{$\eta$\,Ori} is a quintuple system containing several early-type B stars in the suggested configuration (B0.7V + B1.5V) + (B1 Vn + B1 Vn) + B \citep{Maiz2018}. Again, the emission is hardly traceable in the literature and cannot be confirmed by us. We therefore assume that the multiple nature of the system was mistaken for emission.

\paragraph{V355 Per} is classified as early-type contact system by \cite{Avvakumova2013}. As for \object{SX Aur}, we therefore omit this object from our final list of Be stars.

\paragraph{V495 Cen} is an eclipsing interacting binary system consisting of an evolved F-type giant and an early B-type star. The system is currently interacting and the H$\alpha$ emission probably arises from the accretion disk on one of the stars rather than a decretion disk characteristic of classical Be stars \citep{Rosales2018, Rosales2019}.  Moreover, the spectral types of the components are unknown. It is therefore not included in our final sample of early-type Be stars.

\paragraph{V447 Sct} was classified as B0\,Iae. It was found by \citet{Hutchings1973} to be a binary  with a 58\,d period. The authors reported that the H$\alpha$ emission follows an anti-phase motion to the H$\beta$ line, suggesting that there are two MS companions in the system. However, given the luminosity class of the primary, the object cannot be considered a classical Be star and we therefore omit it from our final sample.

\section{Tables}\label{Sec:Appendix:table}

\tiny
\longtab[1]{
\begin{landscape}
\begin{longtable}{lrrrrllll}
\caption{\label{app:table_sample} Stars in the final sample. The first two columns give the star name and the HD number. Columns 3 and 4 give the coordinates of the target, while the next column gives the V-band magnitude. In the column 'SpT' we give the most recent spectral type from literature and in 'SpT Refs' the corresponding spectral type reference. Individual comments on each target are given in the 'Comments' column (see notes for the abbreviations used). The last column for which the table is sorted for indicates the binarity class we assigned it to, where i indicates binaries with confirmed post-MS companion; ii indicates binaries with companions with unknown, uncertain or debated nature; iii indicates Be stars with confirmed MS companions (this class is empty, see Sect.\,\ref{Sec:results}), and iv indicates presumably single stars.}\\ \hline\hline
Name & HD & RA & DEC & V & SpT & SpT Refs & Comments & class \\ 
 &  & J2000 & J2000 & mag &  & & & \\ \hline  
\endfirsthead
\caption{continued.}\\
\hline\hline
Name & HD & RA & DEC & V & SpT & SpT Refs & Comments & class \\ 
 &  & J2000 & J2000 & mag &  & & & \\ \hline  
\endhead
\hline
\endfoot
  V831 Cas &  &  01 47 00.212 &   61 21 23.662 &  11.4 & B1Ve & \cite{Reig1997} & BeXRB \citep{Liu2000} &  i \\
  V615 Cas &         &  02 40 31.664 &   61 13 45.591 &  10.8 &     O9.7(III)e &      \cite{Skiff2014} &                        BeXRB \citep{Massi2018} &    i \\
  V420 Aur &   34921 &  05 22 35.231 &   37 40 33.640 &   7.5 &         B0IVpe &      \cite{Skiff2014} &                          BeXRB \citep{Liu2000} &    i \\
  V725 Tau &  245770 &  05 38 54.575 &   26 18 56.839 &   9.4 &   O9/B0 III/Ve &       \cite{Wang1998} &                       BeXRB \citep{Finger1994} &    i \\
 HD 259440 &  259440 &  06 32 59.257 &   05 48 01.163 &   9.1 &           B0pe &    \cite{Aragona2010} &                      BeXRB \citep{Aragona2010} &    i \\
 ... & & & & & & & & \\
\end{longtable}
\tablefoot{The following abbreviations are used in the comment column: 'App' indicates that the there is a more detailed comment in the appendix \ref{sec:app:comments}. 'Ker19' indicates that \cite{Kervella2019} detect the star in their sample (see Appendix \ref{app:literature}). 'Wang18' means that \cite{Wang2018} included the star in their sample but did not detect the signature of a companion. 'runaway' indicates that the star was classified as runaway candidate in the literature. 'Kle19' indicates that \cite{Klement2019} find an SED turndown indicative of a close binary companion. A full version of this table is available electronically. The first few lines are shown as an example.}
\end{landscape}
}

\longtab[2]{
\begin{landscape}
\begin{longtable}{lrrrrlll}
\caption{\label{App:tableNoBe} Stars that are probably not classical Be stars. The first two columns give the star name and the HD number. Columns 3 and 4 give the coordinates of the target, while the next column gives the V-band magnitude. In the column 'SpT' we give the most recent spectral type from literature and in 'SpT Refs' the corresponding spectral type reference. Individual comments on each target are given in the 'Comments' column (see notes for the abbreviations used.} \\ \hline

Name & HD & RA & DEC & V & SpT & SpT Refs & Comments \\ 
 &  & J2000 & J2000 & mag &  & & \\ \hline  
\endfirsthead
\caption{continued.}\\
\hline\hline
Name & HD & RA & DEC & V & SpT & SpT Refs & Comments \\ 
 &  & J2000 & J2000 & mag &  & & \\ \hline  
\endhead
\hline
\endfoot
    BD+61 105 &         &  00 31 19.201 &   62 25 39.442 &   9.3 &                                  O9 IV &     \cite{Negueruela2004b} &                                                    \\
    BD+63 124 &         &  01 01 12.074 &   63 57 14.405 &  10.9 &                                   B1Ve &         \cite{Jaschek1982} &                                                    \\
    BD+59 334 &         &  01 51 09.313 &   60 26 10.833 &  10.6 &                                   B0 V &         \cite{Hardorp1959} &                                                    \\
    HD 232590 &  232590 &  02 03 48.894 &   55 07 14.523 &   8.6 &                               B1.5IIIe &          \cite{Morgan1955} &                                                    \\
     V355 Per &   13758 &  02 16 04.433 &   57 44 41.674 &   9.1 &                                   B1Ve &         \cite{Jaschek1982} &         App, contact system \citep{Avvakumova2013} \\
     ... & & & & & & & \\
\end{longtable}
\tablefoot{The abbreviation 'em' stands for emission. A full version of this table is available electronically. The first few lines are shown as an example.}
\end{landscape}
}

\longtab[3]{
\begin{landscape}
\begin{longtable}{lrrrrlll}
\caption{\label{App:tableLater} Be stars with spectral types later than B1.5 or unknown spectral type. The first two columns give the star name and the HD number. Columns 3 and 4 give the coordinates of the target, while the next column gives the V-band magnitude. In the column 'SpT' we give the most recent spectral type from literature and in 'SpT Refs' the corresponding spectral type reference. Individual comments on each target are given in the 'Comments' column.} \\ \hline
Name & HD & RA & DEC & V & SpT & SpT Refs & Comments \\ 
 &  & J2000 & J2000 & mag &  & & \\ \hline  
\endfirsthead
\caption{continued.}\\
\hline\hline
Name & HD & RA & DEC & V & SpT & SpT Refs & Comments \\ 
 &  & J2000 & J2000 & mag &  & & \\ \hline  
\endhead
\hline
\endfoot
              EM* AS 3 &         &  00 35 42.334 &   68 42 26.220 &  10.7 &            Be &                           &                                                \\
             BD+60 180 &         &  01 12 11.152 &   61 19 33.175 &   9.3 &     B2.5 IIIe &    \cite{Negueruela2004b} &                                                \\
             BD+56 251 &         &  01 20 50.614 &   57 26 18.925 &  10.3 &          B2e: &        \cite{Kopylov1953} &                                                \\
             EM* AS 16 &         &  01 22 32.561 &   61 32 58.255 &  10.9 &          B3:e &         \cite{Miller1951} &                                                \\
             BD+56 259 &         &  01 23 19.420 &   57 38 51.774 &  10.4 &          B3e: &        \cite{Kopylov1953} & \\
             ... & & & & & & & \\
\end{longtable}
\tablefoot{A full version of this table is available electronically. The first few lines are shown as an example.}
\end{landscape}
}

\end{appendix}
\end{document}